\documentclass[letterpaper,10pt]{article}

\usepackage{amsmath,amssymb,bm,tikz, bbm,
  siunitx,pgfplots,wrapfig,units,float}
\usepackage[skip=1pt]{subcaption}
\usepackage{osameet3} %
\usepackage[font={footnotesize}]{caption}
\usepackage[colorlinks=true,bookmarks=false,citecolor=blue,urlcolor=blue]{hyperref} %
\usepackage{lipsum}
\usepackage[acronym]{glossaries}
\usepackage{todonotes}
\usepackage{marginnote}

\usepackage{tikz}
\usepackage{booktabs}
\usetikzlibrary{fit, positioning, calc, angles}
\usepackage{varwidth}
\newcommand{\figref}[1]{\figurename~\ref{#1}}
\newacronym{bps}{BPS}{blind phase search}
\newacronym{rpn}{RPN}{residual phase noise}
\newacronym{awgn}{AWGN}{additive white Gaussian noise}
\newacronym{gs}{GS}{geometrical shaping}
\newacronym{qam}{QAM}{quadrature amplitude modulation}
\newacronym{snr}{SNR}{signal to noise ratio}
\newacronym{bce}{BCE}{binary cross entropy}
\newacronym{bmi}{BMI}{bitwise mutual information}
\newacronym{gmi}{GMI}{generalized mutual information}
\newacronym{mi}{MI}{mutual information}
\newacronym{e2e}{E2E}{end-to-end}
\newacronym{cpe}{CPE}{carrier phase estimation}
\newacronym{llr}{LLR}{log likelikood ratio}
\newacronym{nn}{NN}{neural network}
\newacronym{tx}{Tx}{transmitter}
\newacronym{rx}{Rx}{receiver}
\newacronym{fec}{FEC}{forward error correction}
\newacronym{bmd}{BMD}{bit-metric decoder}
\newacronym{ff-nn}{FF-NN}{feed-forward neural network}
\newacronym{ber}{BER}{bit error rate}
\newacronym{dsp}{DSP}{digital signal processing}

\newcommand\blfootnote[1]{%
  \begingroup
  \renewcommand\thefootnote{}\footnote{#1}%
  \addtocounter{footnote}{-1}%
  \endgroup
}

\pgfplotsset{
  compat=1.3,
  legend style={
    font=\footnotesize
  },
  every axis label/.append style={
    font=\footnotesize,
    at={(ticklabel cs:0.5)},
    anchor=near ticklabel,
  },
  ticklabel style={font=\footnotesize},
}
\definecolor{ALUColor1}{rgb}{0,0.4470,0.7410}
\definecolor{ALUColor2}{rgb}{0.8500,0.3250,0.0980}
\definecolor{ALUColor3}{rgb}{0.9290,0.6940,0.1250}
\definecolor{ALUColor4}{rgb}{0.4940,0.1840,0.5560}
\definecolor{ALUColor5}{rgb}{0.4660,0.6740,0.1880}
\definecolor{ALUColor6}{rgb}{0.3010,0.7450,0.9330}
\definecolor{ALUColor7}{rgb}{0.6350,0.0780,0.1840}
\definecolor{ALUColor8}{rgb}{0,0.7882,1}
\makeatletter
\DeclareRobustCommand{\rvdots}{%
  \vbox{
    \baselineskip4\p@\lineskiplimit\z@
    \kern-\p@
    \hbox{.}\hbox{.}\hbox{.}
  }}
\makeatother

\begin{document}

\suppressfloats

\title{Geometric Constellation Shaping for Phase-noise Channels Using
  a Differentiable Blind Phase Search}
\vspace*{-2ex}
\author{Andrej Rode, Benedikt Geiger, and Laurent Schmalen}
\address{Communications Engineering Lab (CEL), Karlsruhe Institute of Technology, Kaiserstr. 12, 76131 Karlsruhe, Germany}
\email{\texttt{rode@kit.edu}, \texttt{benedikt.geiger@student.kit.edu}}
\copyrightyear{2022}
\vspace*{-2ex}
\begin{abstract}
  We perform geometric constellation shaping with optimized bit labeling using a binary auto-encoder including a differential blind phase search (BPS). Our approach enables full end-to-end training of optical coherent transceivers taking into account the digital signal processing.%
\end{abstract}\vspace*{+0.6ex}

\glsresetall

\section{Introduction}\vspace*{-1ex}
Modern high-speed coherent optical communication systems rely on feed-forward
\gls*{cpe}, since the high degrees of parallelization and pipelining, required in hardware implementations of coherent receivers, are prohibitive for decision-directed algorithms with feedback. One commonly used state-of-the-art algorithm for \gls*{cpe} is the \gls*{bps}, which was first introduced in
\cite{pfauHardwareEfficientCoherentDigital2009}. %
Classical rectangular \gls*{qam} constellations suffer from a penalty in achievable rate and a gap to the capacity, which can be reduced by constellation shaping techniques. In this paper, we consider geometrical constellation optimization to improve achievable rates. Recently, the use of deep-learning in conjunction with auto-encoders has become a popular method to carry out constellation optimization by taking into account the complete transmission system~\cite{Karanov18},\cite{cammererTrainableCommunicationSystems2020},\cite{Gumus20}. These methods often use simplified models of the transmission, in particular neglecting the influence of \gls*{dsp} algorithms like the \gls*{bps}, as the latter are often not differentiable, a crucial prerequisite to apply deep learning.

Two approaches to
geometrical constellation optimization via deep-learning and auto-encoders for optical systems using a \gls*{bps}
have been presented in
\cite{jovanovicGradientfreeTrainingAutoencoders2021} and \cite{jovanovicEndtoendLearningConstellation2021}. Both
approaches avoid going through the non-differentiable \gls*{bps}; in
\cite{jovanovicGradientfreeTrainingAutoencoders2021}, the authors employ a cubature Kalman filter to approximate the noise distribution at the output of the \gls*{bps} algorithm and in \cite{jovanovicEndtoendLearningConstellation2021}, the authors model the effect of \gls*{bps} via training on \gls*{awgn} and a surrogate \gls*{rpn}. While both approaches show gains compared to regular \gls*{qam}, the resulting constellations are still not optimized for the operation of \gls*{bps} as the end-to-end learning is performed on approximated outputs of the \gls*{bps}. Additionally, the constellations have been
optimized using the \gls*{mi} as target metric. However, most of today's coherent systems employ binary \gls*{fec} schemes with a \gls*{bmd}. The performance of constellations optimized with respect to the \gls*{mi} is suboptimal in this context. 

In this paper, we propose a novel approach to geometric constellation shaping based on end-to-end deep learning with binary auto-encoders~\cite{cammererTrainableCommunicationSystems2020}. To circumvent the issue of the non-differentiable \gls*{bps}, we propose a differentiable version of the \gls*{bps} algorithm that we use during training. The resulting constellations (including a bit labeling) show superior performance in terms of \gls*{bmi}\footnote{The \gls*{bmi} is often referred to as \gls*{gmi} in the optical communications community. We prefer to use the term \gls*{bmi} due to its easier resemblence with the operational meaning.}, in particular in the high \gls*{snr} regime.  Our investigation paves the way for a fully differentiable \gls*{dsp} chain enabling full end-to-end learning of coherent transceivers.\vspace*{-1ex}\blfootnote{This work has received funding from the European Research Council (ERC)
under the European Union’s Horizon 2020 research and innovation programme
(grant agreement No. 101001899).}
\section{System Model of Binary Auto-encoder}\vspace*{-1ex}
We present an approach for a \gls*{gs} constellation optimization for
transmission of an $M$-ary constellation through a laser phase noise affected optical channel where the \gls*{bps} algorithm is used for \gls*{cpe}. Our approach is
based on an auto-encoder with trainable \glspl*{nn} at the
\gls*{tx} and \gls*{rx}. The channel model of the auto-encoder consists of \gls*{awgn}, laser phase noise, and the \gls*{bps} algorithm. The \gls*{bps} algorithm is implemented such that it can be switched between non-differentiable and differentiable mode. With this approach, it can be used in differential mode during training of the \glspl*{nn} and in the original, non-differentiable mode during validation.
\newsavebox\neuralnetwork
\sbox{\neuralnetwork}{%
		\begin{tikzpicture}[
      >=stealth,
      scale=.9,
      every node/.append style={transform shape},
      remember picture,
      ]%
      \tikzset{Source1b/.style={rectangle, draw=black, thick, minimum width=0.05cm, minimum height=1.2cm, rounded corners=0.5mm}}
      \tikzset{Source3/.style={rectangle, draw, thick, minimum width=0.6cm, minimum height=0.45cm, rounded corners=0.5mm}}

      \tikzset{OnehotNode/.style={circle, thick, draw,minimum width=0.1cm}}
      \tikzset{ReLUNode/.style={circle,thick,draw,fill=black!10!white}}
      \tikzset{MZMNode/.style={circle,thick,draw,fill=black!30!white}}
   \node (serial) at (3.6,0) {};
	  \def\offseta{4.5}
	  \def\offsetb{0.5cm}
	  \def\offsetc{15cm}
    \def\k{0}
    \def\ki{1}
        \node [OnehotNode] (S\ki1) at ($(0,0.5)+(0,\k)$) {};
        \node [OnehotNode] (S\ki2) at ($(0,-0.5)+(0,\k)$) {};
		\node at ($(0,0)+(0,\k)$) {$\rvdots$};
		\draw [->] ($(S\ki1)+(-0.4cm,0)$) -- (S\ki1);
		\draw [->] ($(S\ki2)+(-0.4cm,0)$) -- (S\ki2);

		\node [ReLUNode] (H\ki11) at ($(1,1.5)+(0,\k)$) {};
		\node [ReLUNode] (H\ki12) at ($(1,0.5)+(0,\k)$) {};
		\node [ReLUNode] (H\ki13) at ($(1,-0.5)+(0,\k)$) {};
		\node [ReLUNode] (H\ki14) at ($(1,-1.5)+(0,\k)$) {};
		\node at ($(1,0)+(0,\k)$) [anchor=center] {$\rvdots$};

		\foreach\j in {1,2,3,4}{
		  \draw [->] (S\ki1) -- (H\ki1\j); \draw [->] (S\ki2) -- (H\ki1\j);
		};

		\node [ReLUNode] (H\ki21) at ($(2,1.5)+(0,\k)$) {};
		\node [ReLUNode] (H\ki22) at ($(2,0.5)+(0,\k)$) {};
		\node [ReLUNode] (H\ki23) at ($(2,-0.5)+(0,\k)$) {};
		\node [ReLUNode] (H\ki24) at ($(2,-1.5)+(0,\k)$) {};
		\node at ($(2,0)+(0,\k)$) [anchor=center] {$\rvdots$};

		\foreach\j in {1,2,3,4}{
		  \foreach\i in {1,2,3,4}{
		    \draw [->] (H\ki1\j) -- (H\ki2\i);
		  };
		};

		\node [MZMNode] (M\ki1) at ($(3,1)+(0,\k)$) {};
		\node [MZMNode] (M\ki2) at ($(3,0)+(0,\k)$) {};
		\node [MZMNode] (M\ki3) at ($(3,-1)+(0,\k)$) {};
		\draw [<-] ($(serial.west)+(0,\k)+(0,1)$) -- (M\ki1.east);
		\draw [<-] ($(serial.west)+(0,\k)$) -- (M\ki2.east);
		\draw [<-] ($(serial.west)+(0,\k)+(0,-1)$) -- (M\ki3.east);

		\node at ($(3,-0.5)+(0,\k)$) [anchor=center] {$\rvdots$};
		\node at ($(3,+0.5)+(0,\k)$) [anchor=center] {$\rvdots$};
		\foreach\j in {1,2,3,4}{
		  \foreach\i in {1,2,3}{
		    \draw [->] (H\ki2\j) -- (M\ki\i);
		  };
		};
\end{tikzpicture}%
}

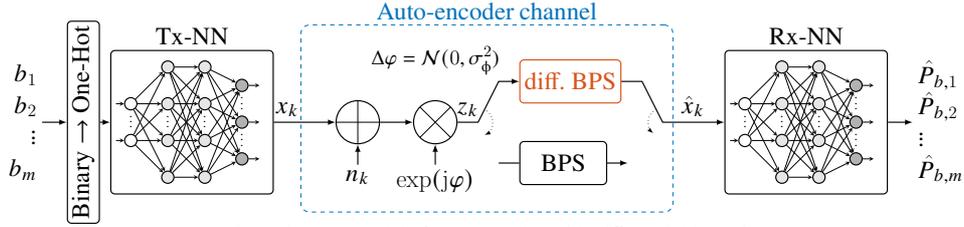
\begin{figure}[t]
    \centering
\tikzset{MUL/.style={draw,circle,append after command={
      [shorten >=\pgflinewidth, shorten <=\pgflinewidth,]
      (\tikzlastnode.north west) edge (\tikzlastnode.south east)
      (\tikzlastnode.north east) edge (\tikzlastnode.south west)
    }
  }
}
\tikzset{ADD/.style={draw,circle,append after command={
      [shorten >=\pgflinewidth, shorten <=\pgflinewidth,]
      (\tikzlastnode.north) edge (\tikzlastnode.south)
      (\tikzlastnode.east) edge (\tikzlastnode.west)
    }
  }
}
\tikzstyle{box}=[rectangle,draw=black, minimum size=7mm, inner sep=2mm, font=\footnotesize, align=center]
\tikzstyle{node}=[circle, draw=black, minimum size=5mm]
\tikzstyle{connection}=[->,>=latex]
\tikzset{%
  cblock/.style    = {draw, thick, rectangle, minimum height = 3em, minimum width = 6em,rounded corners=1mm},
  lblock/.style = {draw,thick,rectangle,minimum height=10em, minimum width=2em,
    rounded corners=1mm},
  operation/.style = {draw, thick, minimum height= 3em, circle},
}
\def\aend{0.25}
	\resizebox{.8\textwidth}{!}{
		\begin{tikzpicture}[font=\LARGE]
		    \node [lblock] (embed) at (0,0) {\rotatebox{90}{Binary $\to$ One-Hot}};

        \node [lblock, right=\aend of embed] (txnet) {\usebox{\neuralnetwork}};

        \node [operation, ADD, right=1.5 of txnet] (awgn_add) {};
        \node [below=0.5 of awgn_add] (awgn_param) {$n_k$};

        \node [operation, MUL, right=0.8 of awgn_add] (phase_mult) {};
        \node [below=0.5 of phase_mult] (pn_param) {$\exp(\mathrm{j}\varphi)$};
		\node [above=0.5 of phase_mult] (pn_param_def) {\Large $\Delta\varphi = \mathcal{N}(0,\sigma_\upphi^2$)};

        \node [coordinate,right=0.5 of phase_mult] (switch_start) {};
        \node [coordinate,right=0.5 of switch_start] (switch_cent) {};
        \node [coordinate,above=1 of switch_cent] (switch_top) {};
        \node [coordinate,below=1 of switch_cent] (switch_bottom) {};
        \pic [draw, angle radius=10, dotted, <-, >=latex] {angle=switch_bottom--switch_start--switch_top};
        \node [cblock, ALUColor2, right=0.5 of switch_top] (diff_bps) {diff. BPS};
        \node [cblock, right=0.5 of switch_bottom] (bps) {BPS};
        \node [coordinate,right = 0.5 of bps] (end_switch_bottom) {};
        \node [coordinate,right = 0.5 of diff_bps] (end_switch_top) {};
        \node [coordinate,below = 1 of end_switch_top] (end_switch_cent) {};
        \node [coordinate,right = 0.5 of end_switch_cent] (end_switch_end) {};
                \pic [draw, angle radius=10, dotted, ->, >=latex] {angle=end_switch_top--end_switch_end--end_switch_bottom};

		    \node [lblock,right=1.5 of end_switch_end] (complexrx) {\usebox{\neuralnetwork}};
		    \node [right=0.6 of complexrx,align=left] (sink) {$\hat{P}_{b,1}$\\ $\hat{P}_{b,2}$\\ $\vdots$\\ $\hat{P}_{b,m}$};
		    \node [left=0.6 of embed,align=right] (source) {$b_1$\\ $b_2$\\ $\vdots$\\ $b_m$};

		    \draw [-latex,thick] (source) -- (embed);
		\draw [-latex,thick] (awgn_param) -- (awgn_add);
		\draw [-latex,thick] (pn_param) -- (phase_mult);
        \draw [-latex,thick] (embed) -- (txnet);
        \draw [-latex,thick] (txnet) -- node[above left]{$x_k$} (awgn_add);
        \draw [-latex,thick] (awgn_add) -- (phase_mult);
        \draw [-latex,thick] (phase_mult) -- node[above]{$z_k$} (switch_start) -- (switch_top) -- (diff_bps);
        \draw [-latex,thick] (switch_bottom) -- (bps) -- (end_switch_bottom);
        \draw [-latex,thick] (diff_bps) -- (end_switch_top) -- (end_switch_end) -- node[above]{$\hat{x}_k$} (complexrx);
        \draw [-latex,thick] (complexrx) -- (sink);

		    \node (rectch) [fit={($(awgn_add.north west)+(-1,2)$) ($(end_switch_end.south east)+(0.25,-2.2)$)}, draw, rounded corners=6pt, dashed, ALUColor1,inner sep=0pt,rectangle] {};
		    \node at (rectch.north) [anchor=north,ALUColor1,above] {Auto-encoder channel};
        \node at (txnet.north) [anchor=north,above] {Tx-NN};
        \node at (complexrx.north) [anchor=north,above] {Rx-NN};
		\end{tikzpicture}}
	\vspace*{-3ex}
  \captionof{figure}{System model of auto-encoder with differentiable \gls*{bps}}
  \label{fig:model}
\vspace*{-4.5ex}
\end{figure}
The system model of the auto-encoder is shown in \figref{fig:model}. The transmitter and receiver of the auto-encoder (modulation order $M$) system are implemented following a binary auto-encoder architecture
\cite{cammererTrainableCommunicationSystems2020},\cite{jones19} and using PyTorch as software framework. %
The \gls*{tx}-\gls*{nn} consists of one fully connected layer, expects one-hot encoded bit vectors $\bm{b}$ of length $m$, with $m = \log_2{M}$ and performs the bit labeling, i.e. assigns a constellation symbol to each bit sequence.
The \gls*{rx}-\gls*{nn} is implemented as a fully connected \gls*{ff-nn} with three layers of width 128 and ReLU as activation function. It is used to estimate the probability $\hat{P}_{b,i} = P(b_i = 1)$ that bit $i \in m$ is a 1. %

The channel model of the auto-encoder between the \gls*{tx} and \gls*{rx}-\gls*{nn} includes \gls*{awgn}, phase noise and a (differentiable) BPS. The \gls*{awgn} manifests in a zero-mean complex normal perturbation $n_k \sim \mathcal{CN}(0, \sigma_n^2)$ which is added to the output of the transmitter, where $\sigma_n^2$ is the noise variance leading to \gls*{snr} $=\frac{1}{\sigma_n^2}$. The phase noise, i.e. laser phase noise, is modeled as a Wiener process with variance $\sigma_{\varphi}^2 =
2\uppi(\Delta f \cdot T_\mathrm{S})$, where $\Delta f$ is the laser linewidth, and $T_\mathrm{S}$ the symbol duration which is the inverse of symbol rate $T_\mathrm{S} = \frac{1}{R_\mathrm{S}}$. The third block of the channel model, the \gls*{bps} is discussed in more detail, because its understanding is essential for this paper.\vspace*{-1.5ex}

\section{Differentiable Blind Phase Search Approximation}\vspace*{-1ex}
To start with, we recapitulate the \gls*{bps} algorithm, as presented in~\cite{pfauHardwareEfficientCoherentDigital2009}: First, the received
complex-valued symbol $z_k$ at time step $k$ is rotated by
$L$ test phases $\bm{\varphi} = (\varphi_0,\ldots,\varphi_{L-1})^T$. %
Second, the squared error $d_{k,b} = |z_k\exp(\mathrm{j}\varphi_b) - \hat{x}_{k,b}|^ 2$ between the closest constellation symbol $\hat{x}_{k,b}$ and $z_k\exp(\mathrm{j}\varphi_b)$, the received symbol rotated by the test phase $\varphi_b$, is calculated for all $L$ test phases ($b \in\{0,\ldots, L-1\}$).
To reduce the impact of noise, a sliding window is used to obtain $s_{k,b} = \sum_{n=-N}^{N}d_{k-n,b}$. Afterwards, the phase is estimated to be $\varphi_{\hat{b}}$, with $\hat{b} = \arg\min_bs_{k,b}$ being the test phase minimizing the sum $s_{k,b}$. Afterwards, a phase unwrap has to be applied to get rid of the phase ambiguities. Finally, the rotated
symbol $\hat{x}_{k,\hat{b}}$ can be computed.

In order to determine the correct test phase, the ``$\arg\min_b$'' operation is applied. Because the $\arg\min_b$ is a non-differentiable operation, the backpropagation algorithm, which is used during deep learning to update the \gls*{tx} \gls*{nn}, cannot be used in combination with the standard \gls*{bps}~\cite{pfauHardwareEfficientCoherentDigital2009}, as it requires a differentiable model. We therefore propose to modify the \gls*{bps} such that it becomes differentiable and can be used during deep learning by the backpropagation algorithm. We use this differentiable \gls*{bps} during training and switch to the standard \gls*{bps} during performance evaluation. To obtain a differential \gls*{bps}, we replace the non-differentiable $\arg\min$ operation by the \emph{softargmin}~\cite{jangCategoricalReparameterizationGumbelSoftmax2017a} operation, which is also commonly referred to as \emph{softmin} in the  Machine Learning community.
The $\mathrm{softmin}(\bm{x})$ function is applied to a vector $\bm{x}$ and returns a vector of identical dimension. The $i$th component of the softmin output is given by\vspace*{-1.4ex}
\begin{align}
  \mathrm{softmin}(x_i) := \left( \mathrm{softmin}(\bm{x}) \right)_i =  \frac{\exp{(-x_i)}}{\sum_j{\exp{(-x_j)}}} 
  \label{eq:softmin}\vspace*{-1.7ex}
\end{align}
where $x_i$ is the $i$th element of $\bm{x}$. To obtain the estimate of the test phase, we compute the the dot product between softmin applied to the vector $\bm{s}_k := (s_{k,0},\ldots, s_{k,L-1})$ and the vector of test phases yielding the differentiable phase estimate $\varphi_{\hat{b},\mathrm{diff.}} = \bm{\varphi}^T\mathrm{softmin}(\bm{s}_k)$.

This approximation works well as long as the phase changes only moderately between $(-\pi, \pi)$. If a phase slip occurs, intermediate results $\varphi_{\hat{b},\mathrm{diff.}}\approx\!\frac{1}{2}\!\left[-\pi\!\cdot\!\mathrm{softmin}(z_k \exp (-\mathrm{j}\pi))\!+\!\pi\!\cdot\!\mathrm{softmin}(z_k \exp (\mathrm{j}\pi))\!\right]\!\approx\!\frac{1}{2}\left(-\frac{1}{2} \pi + \frac{1}{2} \pi \right)\!\approx\!0$ prevent a proper function of the phase unwrap. We hence propose to use a variant of the softmin function, the \emph{softmin with temperature}, which is defined as\vspace{-2ex}
\begin{align}
  \mathrm{softmin}_t(x_i) := \left( \mathrm{softmin}_t(\bm{x}) \right)_i =  \mathrm{softmin}\Big(\frac{x_i}{t}\Big) \label{eq:softmin_temp}\vspace*{-2ex}
\end{align}
For low temperatures ($t \to 0$) the softmin with temperature approximates the $\arg\min$ closely but leads to vanishing gradients. For high temperatures, we get a softer approximation. We use the softmin with temperature during training with a temperature $t$ that decreases steadily from $1$ to $0.001$ during training so that we slowly approach the behavior of the non-differentiable original \gls*{bps}.  %

\vspace*{-0.4ex}
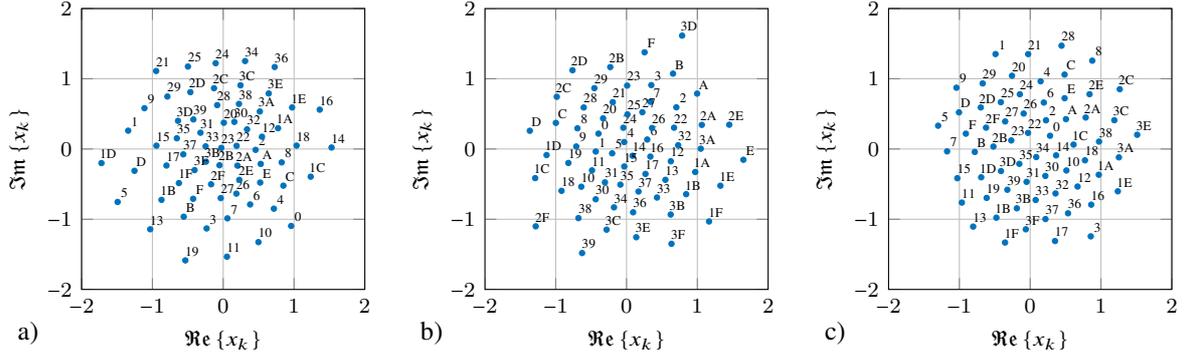
\begin{figure}[!t]
  \centering
\begin{subfigure}{\textwidth/3}\vspace*{-1.5ex}
  \centering
  \begin{tikzpicture}%
    \begin{axis}[
      xlabel={$\mathfrak{Re}\left\{x_k\right\}$},
      ylabel={$\mathfrak{Im}\left\{x_k\right\}$},
      height=\textwidth,
      width=\textwidth,
      xmin=-2,
      xmax=2,
      ymin=-2,
      ymax=2,
      grid=both,
      ]
      \addplot[
      only marks,
      mark=*,
      mark size=1pt,
      color=ALUColor1,
      coordinate style/.from={black,scale=0.5,xshift=5pt},
      nodes near coords,
      point meta=explicit symbolic,
      ]
      table[col sep=tab, meta=label]
      {data/constellation.txt};
    \end{axis}
    \node[below = 3mm,xshift=-0.7cm] {a)};
  \end{tikzpicture}
\end{subfigure}%
\begin{subfigure}{\textwidth/3}\vspace*{-1.5ex}
  \centering
  \begin{tikzpicture}%
    \begin{axis}[
      xlabel={$\mathfrak{Re}\left\{x_k\right\}$},
      ylabel={$\mathfrak{Im}\left\{x_k\right\}$},
      height=\textwidth,
      width=\textwidth,
      xlabel near ticks,
      xmin=-2,
      xmax=2,
      ymin=-2,
      ymax=2,
      grid=both,
      ]
      \addplot[
      only marks,
      mark=*,
      mark size=1pt,
      color=ALUColor1,
      coordinate style/.from={black,scale=0.5,xshift=5pt},
      nodes near coords,
      point meta=explicit symbolic,
      ]
      table[col sep=tab, meta=label]
      {data/constellation_17_600.txt};
    \end{axis}
    \node[below = 3mm,xshift=-0.7cm] {b)};
  \end{tikzpicture}
\end{subfigure}%
\begin{subfigure}{\textwidth/3}\vspace*{-1.5ex}
  \centering
  \begin{tikzpicture}%
    \begin{axis}[
      xlabel={$\mathfrak{Re}\left\{x_k\right\}$},
      ylabel={$\mathfrak{Im}\left\{x_k\right\}$},
      height=\textwidth,
      width=\textwidth,
      xlabel near ticks,
      xmin=-2,
      xmax=2,
      ymin=-2,
      ymax=2,
      grid=both,
      ]
      \addplot[
      only marks,
      mark=*,
      mark size=1pt,
      color=ALUColor1,
      coordinate style/.from={black,scale=0.5,xshift=5pt},
      nodes near coords,
      point meta=explicit symbolic,
      ]
      table[col sep=tab, meta=label]
      {data/constellation_20_300.txt};
    \end{axis}
    \node[below = 3mm,xshift=-0.7cm] {c)};
  \end{tikzpicture}
\end{subfigure}\vspace*{-2ex}
\caption{Geometrically shaped and bit-labeled constellations trained using the differentiable \acrshort*{bps} algorithm at a) \SI{17}{dB} SNR, \SI{100}{kHz} linewidth b) \SI{17}{dB} SNR, \SI{600}{kHz} linewidth c) \SI{20}{dB} SNR, \SI{300}{kHz} linewidth}
\label{fig:constellations}
\vspace*{-5ex}
\end{figure}

\vspace*{-1ex}
\section{Results and Evaluation}\vspace*{-1ex}
We compare the performance of \gls*{gs} constellations learned with two differentiable approximations of the \gls*{bps}: First, the reference model, a binary auto-encoder channel model including only \gls*{awgn} and \gls*{rpn}, which is modeled as a Gaussian distributed phase rotation $\mathcal{N}(0, \sigma_{\mathrm{RPN}}^2)$ and chacterized by the noise variance $\sigma_{\mathrm{RPN}}^2$, as in~\cite{jovanovicGradientfreeTrainingAutoencoders2021}. Second, our approach using a differentiable version of the \gls*{bps} and therefore including a \gls*{awgn}, full laser phase noise and differentiable \gls*{bps} in the auto-encoder channel model. Both approaches are validated using the original, non-differentiable \gls*{bps}. We use the \gls*{bce} during training as loss function to minimize the \gls*{ber} and to maximize 
the \gls*{bmi}~\cite{cammererTrainableCommunicationSystems2020}.
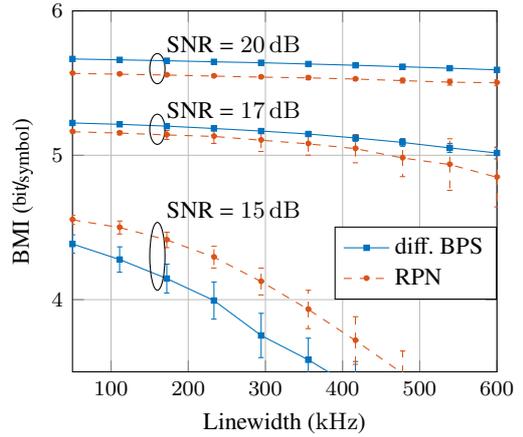
\begin{wrapfigure}[16]{r}{0.45\textwidth}
    \centering
\begin{tikzpicture}
\pgfplotsset{
rpn/.style ={ALUColor2,dashed,mark=*},
bps/.style ={ALUColor1,solid,mark=square*},
}
  \begin{axis}[
    xlabel={Linewidth (\si{kHz})},
    ylabel={BMI (\unitfrac{bit}{symbol})},
    ymin=3.5,
    ymax=6,
    xmin=50,
    xmax=600,
    grid=both,
    width=0.45\textwidth,
    height=0.4\textwidth,
    mark size=1pt,
    legend style={at={(axis cs:600,4)},anchor=south east,nodes={transform shape},font=\small},
    label style={font=\small},
    legend cell align={left},
    ]
    \addplot[
    bps,error bars/.cd,x dir=none, y dir=both, y explicit, 
    ]
    table[x expr={\thisrow{linewidth}/1000}, y=mean, col sep=space, y error=stddev]
    {data/bmi_new_20dB.txt};
    \addplot[
    rpn,error bars/.cd,x dir=none, y dir=both, y explicit
    ]
    table[x expr={\thisrow{linewidth}/1000}, y=mean, col sep=space,y error=stddev]
    {data/bmi_old_20dB.txt};
    \addplot[
    bps,error bars/.cd,x dir=none, y dir=both, y explicit
    ]
    table[x expr={\thisrow{linewidth}/1000}, y=mean, col sep=space,y error=stddev]
    {data/bmi_new_17dB.txt};
    \addplot[
    rpn,error bars/.cd,x dir=none, y dir=both, y explicit
    ]
    table[x expr={\thisrow{linewidth}/1000}, y=mean, col sep=space,y error=stddev]
    {data/bmi_old_17dB.txt};

    \addplot[
    bps,error bars/.cd,x dir=none, y dir=both, y explicit
    ]
    table[x expr={\thisrow{linewidth}/1000}, y=mean, col sep=space,y error=stddev]
    {data/bmi_new_15dB.txt};
        \addplot[
    rpn,error bars/.cd,x dir=none, y dir=both, y explicit
    ]
    table[x expr={\thisrow{linewidth}/1000}, y=mean, col sep=space,y error=stddev]
    {data/bmi_old_15dB.txt};
    \node [font=\small, anchor=south west] at (axis cs:160,5.65) {\gls*{snr} $=\SI{20}{dB}$};
    \node [font=\small, anchor=south west] at (axis cs:160,5.19){\gls*{snr} $=\SI{17}{dB}$};
    \node [font=\small, anchor=south west] at (axis cs:160,4.5){\gls*{snr} $=\SI{15}{dB}$};
  \legend{diff. BPS, RPN};
  
  \draw (axis cs:160,5.6) ellipse (0.1cm and 0.2cm);
  \draw (axis cs:160,5.18) ellipse (0.1cm and 0.2cm);
  
  \draw (axis cs:160,4.3) ellipse (0.1cm and 0.45cm);
  \end{axis}

\end{tikzpicture}\vspace*{-2.5ex}
\caption{Validation results with non-differentiable BPS}\vspace*{-3ex}
\label{fig:performance_comparison}%
\end{wrapfigure}
The learned posteriors of the \gls*{rx}-\gls*{nn} are used to estimate the \gls*{bmi}~\cite{jovanovicGradientfreeTrainingAutoencoders2021}. The training was performed for $M=64$ with hyperparameters as defined previously. %
For the training of both models a \gls*{snr} of \SI{17}{dB} has been selected, and for our approach, a fixed linewidth of \SI{100}{kHz} at \SI{32}{Gbaud} symbol rate, $L = 60$ test phases in $(-\pi, \pi)$, and a sliding window length of $N = 60$ was chosen, while the reference constellation was trained on the fixed surrogate \gls*{rpn} channel with $\sigma_{\mathrm{RPN}} = 0.005$ as in~\cite{jovanovicEndtoendLearningConstellation2021}. 

The validation between both approaches is performed across three \gls*{snr} levels and on ten equally spaced linewidths between \SI{50}{kHz} and \SI{600}{kHz}, while the other parameters are the same as during training. %
We simulated 100 validation runs for each setting on both constellations with $10^5$ symbols per run. In the plot, we provide the mean together with error bars indicating the standard deviation. The achieved \gls*{bmi} of the trained models, training channel is used as label, is shown in \figref{fig:performance_comparison}. %
In the low \gls*{snr} regime, the auto-encoder channel model with \gls*{awgn} and \gls*{rpn} outperforms our approach. Understanding this behavior is part of our ongoing investigations. 
At high \glspl*{snr}, the auto-encoder channel model including the differentiable \gls*{bps} outperforms the reference auto-encoder model including only \gls*{awgn} and \gls*{rpn}. Additionally, our approach is more robust (lower variance during validation) which can be observed by comparing the results at an \gls*{snr} of 17\,dB, especially for high laser linewidths.
The constellation trained on the differentiable \gls*{bps} achieves a gain of
$\SI{0.1}{\unitfrac{bit}{symbol}}$ across the full range of linewidths for a \gls*{snr} of 20\,dB. Hence, an accurate modeling of the channel including \gls*{dsp} algorithms like the \gls*{bps} is important to harness the full potential of \gls*{gs} and increase the achievable rate. 

The learned 64-ary constellations with the corresponding bit labels (given as hexadecimal numbers) are shown in \figref{fig:constellations}. During training, the constellations develop mostly Gray-coded bit labels and a slightly asymmetric shape. 
This can be observed for example at the lower left of \figref{fig:constellations}-a) with symbols $5$, $13$, $19$, which have a larger spacing to neighbouring
symbols compared to anywhere else in the constellation. Interestingly, this slight asymmetricity improves the \gls*{bps} performance,
by providing constellation points the \gls*{bps} algorithm can latch-on to.\vspace*{-1.5ex}

\section{Conclusion}\vspace*{-1ex}
With the proposed differentiable \gls*{bps}, we obtain geometrically shaped constellations that are well suited for use with binary codes and outperform constellations that have been obtained with surrogate residual phase noise, in particular in the high \gls*{snr} regime due to the use of a more exact model of the transmission chain. This work shows how part of the receiver can be replaced by its differentiable counter-parts and paves the way for future end-to-end-optimization of complete optical transceivers.

\end{document}